\documentclass[useAMS,usenatbib]{mn2e}
\usepackage{epsfig,lscape} 
\usepackage{natbib}
\usepackage{amsmath}

\title[Inside-out growth or inside-out quenching?]
{Inside-out growth or inside-out quenching? clues from colour gradients of local galaxies}

\author[J. Lian et al.]
{Jianhui Lian,$^{1,2}$\thanks{ljhhw@mail.ustc.edu.cn (J. Lian); renbin@pa.uky.edu (R. Yan);
Blanton@physics.nyu.edu (M. Blanton); xkong@ustc.edu.cn(X. Kong)}
Renbin Yan,$^{1}$ Michael Blanton,$^{3}$ Xu Kong,$^{2}$\\
$^{1}$Department of Physics and Astronomy, University of Kentucky, Lexington, Kentucky 40506, U.S.\\
$^{2}$CAS Key Laboratory for Research in Galaxies and Cosmology, Department of Astronomy,
University of Science and Technology
of China, Hefei, Anhui 230026, China\\
$^{3}$Center for Cosmology and Particle Physics, Department of Physics, New York University, NY 10003, U.S.}

\begin{document}
\maketitle

\begin{abstract}
{We constrain the spatial gradient of star formation history within galaxies using the colour gradients in NUV$-u$ and $u-i$ for a local spatially-resolved galaxy sample.}
By splitting each galaxy into an inner and an outer part, 
we {find} that most galaxies {show} negative gradients in these two colours.
{We first rule out dust extinction gradient and metallicity gradient as the dominant source for the colour gradient.}
{Then using stellar population models, we explore variations in star formation history to explain the colour gradients. As shown by our earlier work, a two-phase SFH consisting of an early secular evolution (growth) phase and a subsequent rapid evolution (quenching) phase is necessary to explain the observed colour distributions among galaxies.
We explore two different inside-out growth models and two different inside-out quenching models by varying parameters of the SFH between inner and outer regions of galaxies. Two of the models can explain the observed range of colour gradients in NUV$-u$ and $u-i$ colours. We further distinguish them using an additional constraint provided by the $u-i$ colour gradient distribution, under the assumption of constant galaxy formation rate and a common SFH followed by most galaxies. We {find} the best model is an inside-out growth model in which the inner region has a shorter e-folding time scale in the growth phase than the outer region.   
More spatially resolved ultraviolet (UV) observations are needed to improve the significance of the result.} 
\end{abstract}


\section{Introduction}
The cosmic star formation history (SFH) of the universe is well established \citep{madau2014}. However, the SFH of
individual galaxies which contribute to the cosmic SFH is not  {fully understood} yet.
For galaxies on the  {stellar mass-star formation rate (SFR, i.e. star formation main-sequence)} relation, their SFH is relatively well determined  {and can be approximated by a single phase of exponential decline} \citep{noeske2007}. 
{Generally, the SFR of star-forming galaxies declines slowly with cosmic time with an exponential timescale
typically longer than 10 Gyr \citep{noeske2007}.} 
{Meanwhile, more} massive galaxies tend to form 
at an earlier time and have SFR decreasing slightly faster.
Off the main-sequence relation, the situation becomes complicated and unclear. Many previous works 
suggest that galaxies just leaving the star-forming main-sequence are experiencing a
rapid decline of SFR (\citealt{faber2007,martin2007,balogh2011,lian2016}, hereafter
Paper I) {with a timescale of the order of 1 Gyr \citep{scha2014} or even lower \citep{lian2016}}. 
{This rapid cessation process of SFR is distinctive} from the secular evolution in the star-forming stage,
and {is often} referred {to as a} `quenching' process. 
This quenching process is necessary {in galaxy evolution models} to explain the galaxy distribution in NUV$-$optical colour-colour diagrams
\citep{scha2014,lian2016} and the {buildup} of red-sequence from $z=1$ to $z=0$ 
\citep{bell2004,blanton2006,faber2007,moustakas2013}. 
In Paper I, we explored two key properties of the quenching process, the evolution speed 
during quenching and the quenching rate among star-forming galaxies, by analyzing  {the} galaxy distribution 
in NUV$-u$ versus $u-i$ colour diagram.

{Aside from the unclear} integrated SFH of galaxies, 
{how the SFH of galaxies varies internally is still poorly understood.} 
{One direct probe of the internal variation of SFH is the spatial distribution of colours, i.e. the colour gradient.}
Negative colour {gradients have} been found to be prevalent in nearby galaxies 
(e.g. \citealt{peletier1990}). 
However, the physical origin of this colour gradient is {not fully understood and is under debate. Spatial variations of galaxy properties other than the SFH are in principle able to produce the observed negative colour gradient, such as } 
dust extinction \citep{liu2016} and {stellar} metallicity 
\citep{wu2005,tortora2010}. 
{Therefore, it is important to understand the role played by these parameters and SFH in the colour gradient before we use this colour gradient to study the internal variation of SFH.}
{After accounting for the contribution by dust extinction and stellar metallicity, the observed colour gradient could be used as a powerful tool to investigate the spatial variations of SFH in a galaxy. 
However, interpreting the colour gradient as a gradient in SFH is not straightforward since the SFH is usually complicated.}

{In addition, the phrase people use to describe the spatial variation in SFH within a galaxy is confusing the situation. Both `inside-out growth' and `inside-out quenching' can be found in the literature, most of the time without clear definitions. Some would even consider these two phrases as describing the same phenomenon. At the same time, the same phrase has been used to refer to different observation by different authors. For example, `inside-out growth' has been used to refer to the growth of the disk truncation radius or scale radius \citep{azzollini2008}, or the gradient in specific SFR \citep{wang2011}.

{In paper I, we demonstrated that a two-phase SFH is necessary to explain the colour distribution of galaxies in NUV$-u$ vs. $u-i$ diagram. One phase is a secular evolution phase with long timescales, which we define as the `growth' phase. It is followed by a second phase with a rapid decline of SFR on short timescales, which we define as the `quenching' phase. This allows us to clearly distinguish growth from quenching. 
In this work, we use the same SFH as in Paper I. To constrain the spatial variations in SFH, we use stellar population synthesis model with various spatial variations in SFH to reproduce the observed negative colour gradients. 
The models with spatial variation in SFH at the growth phase or the quenching phase are classified as 
`inside-out growth' or `inside-out quenching' models, respectively. 
{As shown in Paper I}, 
NUV$-$optical colours are sensitive to the quenching process and could be used to constrain the quenching properties.}
Thanks to the large sky area surveys of {\sl GALEX} in UV and SDSS in optical bands,  
the NUV$-$optical colour gradient is accessible for a statistically significant sample \citep{liu2016,pan2016}.
Throughout this paper, we adopt the cosmological parameters with $H_0=70\, {\rm km s^{-1} Mpc}^{-1}$, $\Omega_{\Lambda}=0.73$ 
and $\Omega_{\rm m}=0.27$. All magnitudes in this paper are given in the AB photometric system.

\section{Sample selection}
We select a local galaxy sample from NASA Sloan Atlas {(NSA)} catalogue \citep{blanton2011}\footnote{http://www.nsatlas.org/data}, which provides SDSS and {\em GALEX} images, consistently measured photometry at UV and optical bands, and rest-frame absolute magnitude at these bands after correcting for Galactic extinction and k-correction.  
{It also provides axis ratio and position angle measurements determined from Stokes parameters at 90\% light radius.}
{In this work, we consider the regions within 0.3 effective radius ($r_{\rm e}$) of a galaxy as its inner part while the regions with 
$0.5r_{\rm e}<r<r_{\rm e}$ as the outer part. We define these regions by 
elliptical shapes with identical axis ratios and position angles as those measured for the overall galaxy, which is given by the NSA catalogue. 
The gap between the two spatial parts is used to minimize the potential light contamination between them.}
{We require the inner part ($r< 0.3 r_{\rm e}$) to cover}
{at least} one {\em GALEX} PSF (FWHM of 
5.3$''$ at NUV band) along its minor axis. 
{Given the lower limit of the axis ratio we allow, this means a galaxy needs to have a effective radius $r_{\rm e}>12''$ to be selected.}
With such a stringent requirement, {we could only select nearby massive galaxies to ensure a reasonable mass completeness. Figure 1 shows the distribution of galaxies with $z<0.01$ in the mass-$r_{\rm e}$ diagram.
The {dashed horizontal} line indicates the size limit required to resolve the inner part of galaxies. To achieve a relatively high completeness 
(i.e. high fraction of galaxies selected by the size criteria), a high lower limit on mass needs to be set due to the positive mass-size relation.}
To minimize intrinsic extinction, we focus on face-on galaxies with axis ratios ${\rm b/a}>0.7$. 
To summarize,
the selection criteria {for our galaxy sample} are: (1) stellar mass $M_*> 10^{9.5} {\rm M_{\odot}}$; 
(2) redshift $z<0.01$; (3) effective radius $r_{\rm e}>12''$; (4) axis ratio ${\rm b/a}>0.7$.
Finally, {76\% of galaxies with mass above the threshold (229)} objects are selected as our local spatially-resolved galaxy sample.
{Among them, 102 galaxies with active star formation activity are detected in the WISE W3 ($12\mu$m) band.
} 

\begin{figure}
	\centering
	\includegraphics[width=9cm]{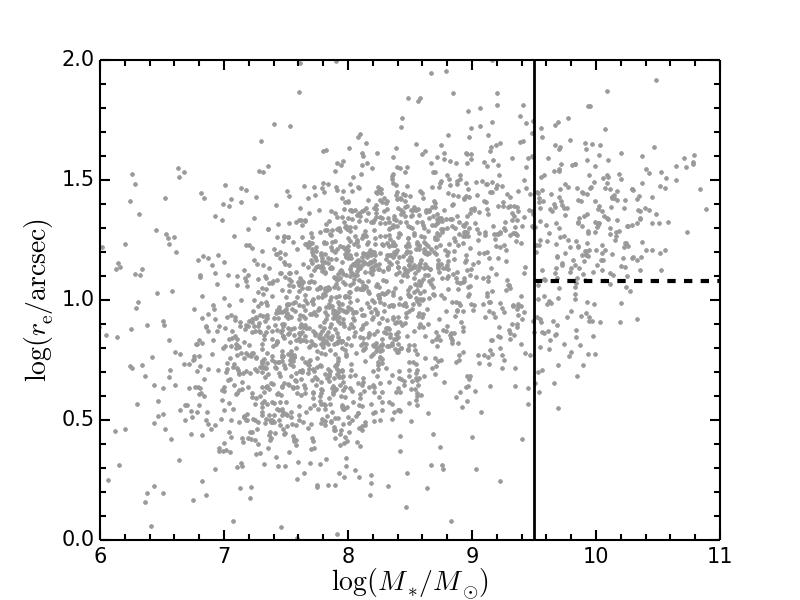}
	\caption{Mass-size relation of galaxies with redshift $z<0.01$. {The} Dashed horizontal line indicates the size limit required to resolve the inner part of galaxies while {the solid} vertical line denotes the mass limit to ensure an acceptable mass completeness.  
	}
	\label{figure1}
\end{figure}

\section{Data reduction}
For {our} galaxy sample, we use the {\em GALEX} and SDSS images provided by NSA \citep{blanton2011}. 
WISE images, reduced and released by \citet{lang2014}\footnote{http://unwise.me/}, are also used 
to constrain the dust extinction in galaxies. 
{To perform photometry in a consistent way, images in various {wavelength} bands need to be 
recentered to the same coordinate and converted to the same resolution level.} 
{We convolve} the optical images with a gaussian kernel to match the 
PSF of UV images and then rebin {them} to the same pixel size of 1.5$''$. 
To obtain the W3$-$NUV colour, we generate another set of UV images that are converted to the same PSF (6$''$) and pixel 
size (2.5$''$) of WISE images.  

For each {galaxy} in our sample, we measure the magnitudes of its inner and outer parts by {summing the} flux in 
the corresponding elliptical annulus.
Since the k-correction is rather small for such nearby objects, with typical values less than 0.05 mag, we 
 {apply} the correction factors 
derived from integrated {magnitudes} in the NSA catalog {to} the inner and outer parts.
The correction factor of foreground Galactic extinction is also taken from the NSA catalog 
{and assumed to be identical for the two parts of the same galaxy.}
{As the} measurements {are obtained} at fixed {effective} radius, we refer to the {spatial} variations of properties between the two parts
as {radial} gradients.

\section{colour gradients and their physical origin}
\subsection{Colour distribution}
{In paper I, we discussed the NUV$-u$ colour {distribution} of galaxies in detail. In general, the distribution of galaxies in colour space is bimodal. Blue star-forming and red-sequence galaxy populations are separated by a wide low density region called `green valley'. The boundary of these three regions in colour space can be identified by significant number density drops, suggesting distinctive evolution stages of galaxies in these regions. Moreover, in order to reproduce the galaxy distribution in colour-colour space, we have to use a two-phase SFH with the second phase having much faster declining SFR. }
{Figure 2} shows the {distribution} of the inner and outer parts of galaxies in {the} NUV$-u$ and $u-i$ colour {space}, denoted by dashed red and solid blue lines, respectively. 
{Interestingly, similar to the distribution of integrated colours in Paper I, the NUV$-u$ colour distribution of the two parts of galaxies are both bimodal. These common bimodal feature strongly suggests the inner and outer parts experience very similar two-phase SFH to the one adopted in Paper I. }
{However,} the inner parts are {systematically} redder than the outer parts in both NUV$-u$ and $u-i$ colour. 
{In NUV$-u$, the median of the colour difference is 0.23 mag; in $u-i$, the median of the colour difference is 0.19 mag}. 
{This negative colour offset} is consistent with previous work 
\citep{pan2014,pan2016} where the authors found similar negative gradient in the NUV$-$optical colours in local galaxies.
{Nevertheless, there} is a small fraction of  galaxies having blue star-forming outer part but red quiescent inner part.}

\subsection{Gradient in dust extinction}
{To investigate the origin of the negative colour gradient, we first explore the role played by the galaxy internal dust extinction.}
{We} obtain rough estimates {of the} dust extinction of the inner and outer parts using the  
IRX-$A_{\rm UV}$ relation \citep{meuer1999,kong2004,buat2005,hao2011}.
{The }IRX index is defined as a  {flux} ratio between the infrared (TIR) and ultraviolet (UV) ($L({\rm TIR})/L({\rm UV})$)  {bands}.
This index was proposed based on energy budget balance and has been widely-used as a robust tracer of dust extinction.
{The} NUV$-$W3 colour  {can be used} as an approximation to the IRX index.
The left panel in  {Figure 3} shows the distribution of NUV$-$W3 colour of the inner and outer parts, 
{with median value of 4.37$\pm0.92$ and 3.99$\pm0.82$ mag, respectively.} 
{It can be seen that} the inner parts  {are slightly} redder {than the outer parts in the} NUV$-$W3 colour, indicating that they {suffer more dust extinction}. 

{To estimate the IRX index, we obtain the IR luminosity using the luminosity in WISE W3 band and the conversion from 
luminosity in a single mid-IR band to total IR  {(TIR)} luminosity proposed by \citet{rieke2009}.}
{The IR luminosity calibration by \citet{rieke2009} for the mid-infrared $12\mu$m band is based on the photometry in {\sl IRAS} bands 
as ${\rm log}(L({\rm TIR})) = (-0.947\pm0.324)+(1.197\pm0.034){\rm log}(L(12))$. Because of the good agreement between the photometry in WISE W3 and {\sl IRAS} 12$\mu$m band \citep{lian2014}, we 
use the calibration in \citet{rieke2009} directly to convert the luminosity in WISE W3 band to the total IR luminosity.}
{Dividing the IR luminosity by the UV luminosity derived from the UV images}, we obtain the IRX index for the two parts of galaxies. 
{According to \citet{buat2005}, the dust extinction at NUV band can be derived using the IRX index by the calibration 
$A_{\rm NUV} = -0.0495x^3+0.4718x^2+0.8998x+0.2269$ where $x$ is the IRX index.}
{Using this calibration,} we acquire the extinction $A_{\rm NUV}$ as shown in the 
right panel of {Figure 3 with median values of $4.30\pm1.00$ and $4.23\pm0.86$ for the inner and outer parts, respectively}. The median offset is slight {and not statistically significant, meaning} 
a negligible negative gradient in dust extinction. Furthermore, assuming the extinction law of \citet{cardelli1989},
we calculate the colour gradient introduced by this slight gradient in extinction, which is 
0.029 in $u-i$ and 0.043 in $NUV-u$, about 10 times lower than the observed colour gradients.
Therefore, the dust extinction can be ruled out as the main {driver} of {the} observed colour gradient.

{It should be noted} that the dust extinctions derived above are just rough estimates and may not be precise 
enough to be used to correct the fluxes in UV and optical bands.
{Comparing} to the intrinsic dust extinction estimates based on stellar continuum fitting 
\citep{ossi2011}, our results overestimate the extinction by a factor of 2. There could be 
several possible reasons for this discrepancy. In deriving {the total} IR luminosity, we ignored
the potential difference in {\sl IRAS} 12$\mu$m and WISE W3 photometry.
Another problem is that the {total IR luminosity} calibration may not be reliable {in the low infrared luminosity region}. 
This calibration is valid when $L(12\mu{\rm m})>10^{8.5}L_{\odot}$ and will {potentially}
overestimate the IR luminosity for objects with fainter $L(12\mu$m). {The} majority of the inner and outer parts 
of galaxies in our
sample have $L(12\mu{\rm m})$ {within} $10^{7.5}-10^{8.5}L_{\odot}$. 
{Lastly}, the IRX-$A_{\rm NUV}$ relation in \citet{buat2005} is derived for {a}
stellar population model with {a chosen} SFH {that} may not be {representative to the galaxies in our sample.}
Although these problems could affect the absolute value of 
dust extinction, {the effect should be similar {in} the inner and outer parts of galaxies and the dust extinction gradient estimates 
should not be significantly biased.}

\begin{figure*}
\centering
\includegraphics[width=15cm]{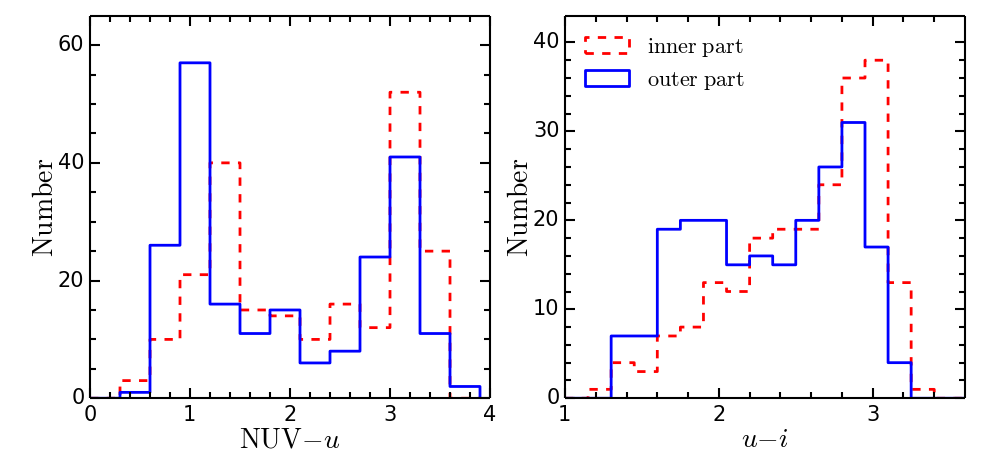}
\caption{Number density profile of the inner and outer parts of galaxies in NUV$-u$ (left panel) and $u-i$ colour (right panel) space. 
}
\label{figure2}
\end{figure*}

\begin{figure*}
\centering
\includegraphics[width=15cm]{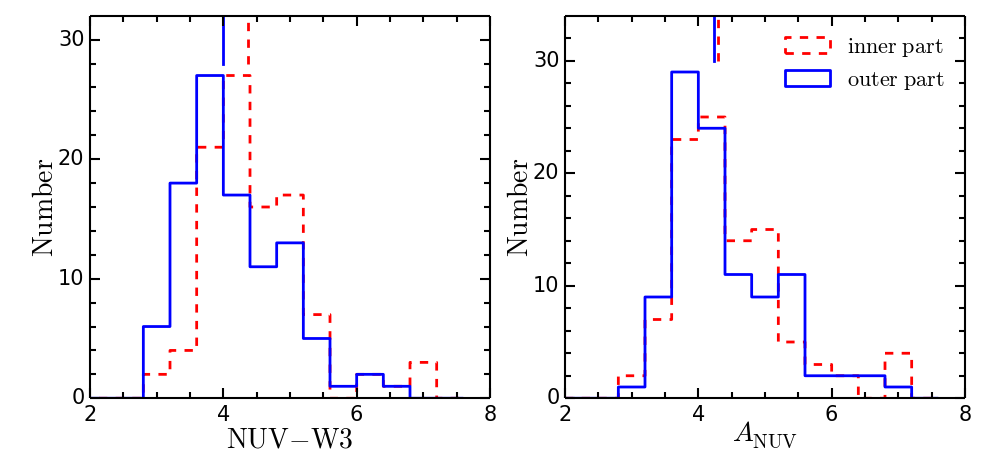}
\caption{Left panel: NUV$-$W3 colour distribution of the inner and outer parts of galaxies. Right panel: 
Extinction in NUV band A$_{\rm NUV}$ distribution of the two parts. Vertical lines indicate the median value.
}
\label{figure3}
\end{figure*}

\subsection{Gradient in star formation history}
\subsubsection{Stellar population model}
In Paper I, we {used a stellar population synthesis model and} assumed a SFH with a two-phase exponential decline. 
{The model can successfully reproduce the curved locus populated by galaxies in the colour-colour diagram and the NUV$-u$ colour distribution.} 
In this work we {use the stellar population model with a similar} two-phase SFH to {reproduce} the colours of the inner and outer parts 
of galaxies.
Here we briefly introduce the parameters in the model. For more detailed description please see \textsection4.1 in Paper I.
There are four parameters in the two-phase SFH, {galaxy} formation time $t_{\rm f}$, {initial e-folding time at the growth phase $\tau_{\rm g}$,} quenching start time $t_{\rm q}$, and e-folding time $\tau_{\rm q}$ {at the quenching phase}.
{The first two parameters are for the growth phase and the last two are for the quenching phase.}
{The left panel of Figure 4}
{shows a schematic plot of} the two-phase SFH and {illustrates the four free} parameters.
The growth phase and quenching phase of the SFH are denoted by cyan and orange lines, respectively.
We generate model {spectral energy distribution (SED)} using \citet{bc03} model spectra with a \citet{chabrier2003} {stellar} initial mass function.
Following {the work by} \citet{scha2014} we {exclude} contribution by stellar population younger than 3 Myr because they tend to be surrounded by optically thick clouds
and are not visible in the UV and optical.
{For the stellar metallicity in the model, we first adopt solar abundance ($Z=0.02$).}

As {we found} in Paper I, NUV$-u$ and $u-i$ colours {can provide constraints on} the parameters 
of the two-phase SFH.
Generally, $u-i$ traces the average age of the stellar population while NUV$-u$ is 
more sensitive to the strength of {the recent} star formation activity. Combining these two colours 
{could constrain} how fast the SFR {declines (i.e. the e-folding time) during the quenching phase}.  
In the right panel {of Figure 4}, the black contour represents the distribution of galaxy sample in Paper I.
{It can be seen that the model reproduced the curved locus in the colour-colour diagram. For more discussion about the 
{effects} caused by different e-folding {times} at the growth and at the quenching phase, please refer to Figure 2 in Paper I. }

{It has been found that star-forming galaxies typically have {a} negative 
stellar metallicity gradient \citep{gonzalez2015,goddard2017a} which may contribute to the observed negative colour gradient \citep{wu2005,tortora2010}.
To study the origin of {the} colour gradient, it is necessary to investigate the role played by the metallicity gradient. 
Therefore, we also generate another model with metallicity set to be $-0.7$ dex
to investigate the potential colour gradient introduced by the negative metallicity gradient.}
In Figure 4, the squares connected by solid line show the solar abundance model, while the circles connected by dashed line show the sub-solar abundance model. The colour scheme is the same as the left panel to indicate the growth and quenching phase.  
It can be seen that the metal-poor model shows slightly bluer colour compared to the model with solar-abundance at the 
growth phase and much more significant colour difference at the quenching phase.
{During the growth phase, the} maximum colour offset achieved at the growth phase by this 0.7 dex difference in stellar metallicity is $\sim0.2$ mag in $u-i$ {and less than 0.1 mag in NUV$-u$. {The variation in colour} caused by {the} stellar metallicity gradient is} much smaller than the observed colour gradients, {especially in NUV$-u$}. 
Moreover, the stellar metallicity gradient in star-forming galaxies is much flatter than $-0.7$ dex/$r_{\rm e}$ with a typical gradient of $-0.1$ dex/$r_{\rm e}$ \citep{goddard2017b}. 
{Although the colour difference caused by variation in stellar metallicity {tends} to be larger after the galaxy is quenched, the stellar metallicity gradient in passive galaxies is flatter than {for} star-forming galaxies \citep{goddard2017b}.}
Therefore, we conclude that the stellar metallicity gradient cannot be the driver of the observed colour gradient.

\begin{figure*}
\centering
\includegraphics[width=15cm]{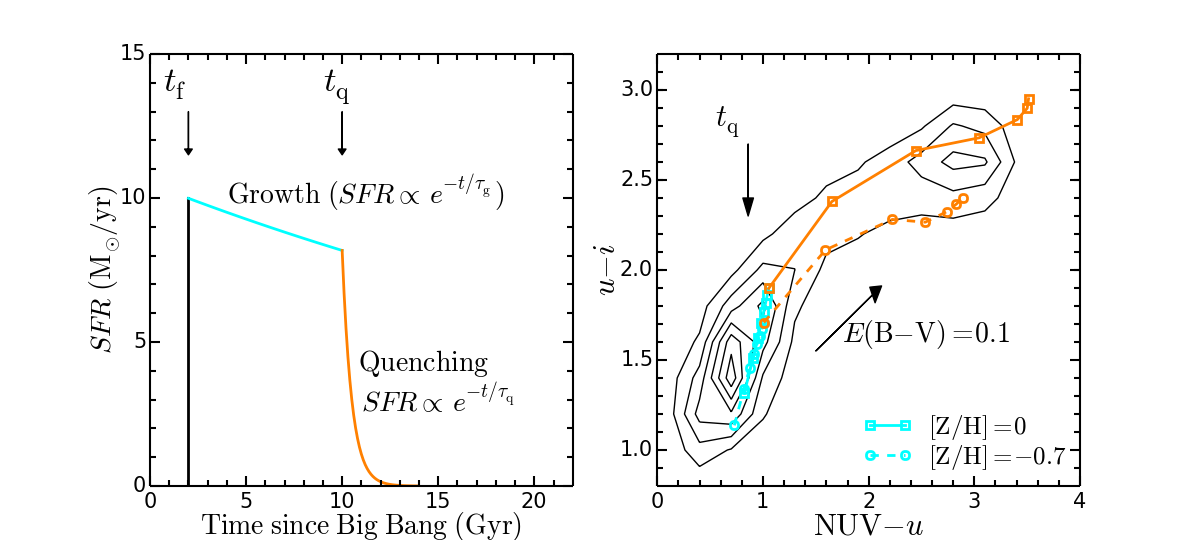}
\caption{Schematic plot to show the two-phase SFH adopted in stellar population model. 
The left panel shows the two-phase SFH definition and the four parameters that 
characterize the SFH. Cyan and orange lines denote the growth and quenching phase of the SFH, respectively.
The right panel shows this model's predictions for
NUV$-u$ v.s. $u-i$ colours. The black contour represents the distribution of our galaxy sample. 
Squares represent the model with metallicity fixed to the solar value while the circles are the model with sub-solar metallicity [Z/H] of $-0.7$ dex. {The points on the model tracks denote the age of the model from 1 Gyr to 14 Gyr with a interval of 1 Gyr.}
The black arrow indicates the reddening caused by extinction with colour excess E(B$-$V)=0.1.
}
\label{figure4}
\end{figure*}

\subsubsection{Gradients in SFH}
In {previous sections} we found systematically negative colour gradients in our galaxy sample and excluded the 
gradient in dust extinction {and stellar metallicity} as {the dominant drivers for the colour gradient}.
{In this subsection}, we further investigate what gradient in SFH could  {produce} the
observed colour  {gradients in most} galaxies.
{Figure 5 shows the NUV$-u$ and $u-i$ colour gradients as a function of NUV$-u$ colour of the outer part in the first two columns and 
the relation between the two colour gradients in the third column.}
{The data points represent individual galaxies and are identical in different rows of Figure 5.}
{According to the analysis in Paper I, the colour information of quiescent galaxies is not critical to constrain the SFH. 
In this figure, we exclude 89 galaxies with quiescent red outer parts ((NUV$-u)_{\rm outer}>2.7$).
This colour cut for quiescent galaxies is based on a clear number density drop at NUV$-u\sim2.7$ found in Paper I.}

It can be seen that most galaxies in our sample show a negative gradient in both NUV$-u$ and $u-i$ colour. 
This is consistent with the systematic  {shifts of colour distributions shown in Figure 2}. 
The colour gradient in $u-i$  {seems to decrease} as the outer part becomes redder in NUV$-u$, 
while  {no clear trend but large scatter can be found in the gradient in NUV$-u$.}
{Interestingly, a minor fraction (30\%) of galaxies have positive colour {gradients} either in NUV$-u$ or $u-i$.
These positive colour gradients may be due to gas infall in the center that triggers star formation or stripping of the gas in the outer parts, {suppressing} the star formation. Investigating the origin of the positive colour gradient is not the goal of this work and we focus on the remaining 103 non-quiescent galaxies with negative colour gradient hereafter.}
 
To investigate whether and how  {a} gradient in SFH  {can reproduce the observed negative colour gradient, we generate four models of spatially-varying SFH. In each model, we have two SFHs, one for the inner region and one for the outer region. As mentioned in Sec 4.3.1, each SFH is described by four parameters. For simplicity, the two SFHs in each model share the same values for three of the four parameters and differ in only one of them. We refer to the varying parameter as the `gradient' parameter. Each model explores a different `gradient parameter' and is shown in each row of Figure 5. 
The fourth column of Figure 5 shows the SFHs adopted for the corresponding model. 
{In contrast to the rest of the figure, in the first row we are varying the SFH of the outer part, while keeping the inner part fixed.} The black line represents the SFH adopted for the outer part except for the model in the first row where the black line indicates the inner part. The other three {coloured} lines show three possible SFHs for the other part with the gradient parameter taking on different values. For clarity, a different normalization is adopted for each SFH.}
{It should be noted that much more sophisticated models with more than one parameter varying might be less likely, but {could} to explain the colour gradient as well.} 

Initially, there are  {six} free parameters in the  {each model, four parameters for one SFH, one additional parameter for the other SFH which has a different value, and dust extinction.} For simplicity, the formation time of the inner part is fixed to be zero. 
{The dust extinction is assumed to be $E({\rm B-V})=0.1$ to match the NUV$-u$ colour of galaxies in the horizontal direction.}
{The number of free parameters is reduced to four.}
In this work, we do not intend to fit the data points one by one but 
explore the possible colour gradients that could be produced by the models. 
Therefore, we only show two or three models with a sparse grid in the `gradient parameter'.
{When they are not used as a gradient parameter, $t_{\rm f}$, $t_{\rm q}$, $\tau_{\rm g}$, and $\tau_{\rm q}$ are fixed to 0 Gyr, 10 Gyr, 40 Gyr and 0.8 Gyr, respectively.}

\begin{figure*}
\centering
\includegraphics[width=18cm,viewport=160 130 1760 1700,clip]{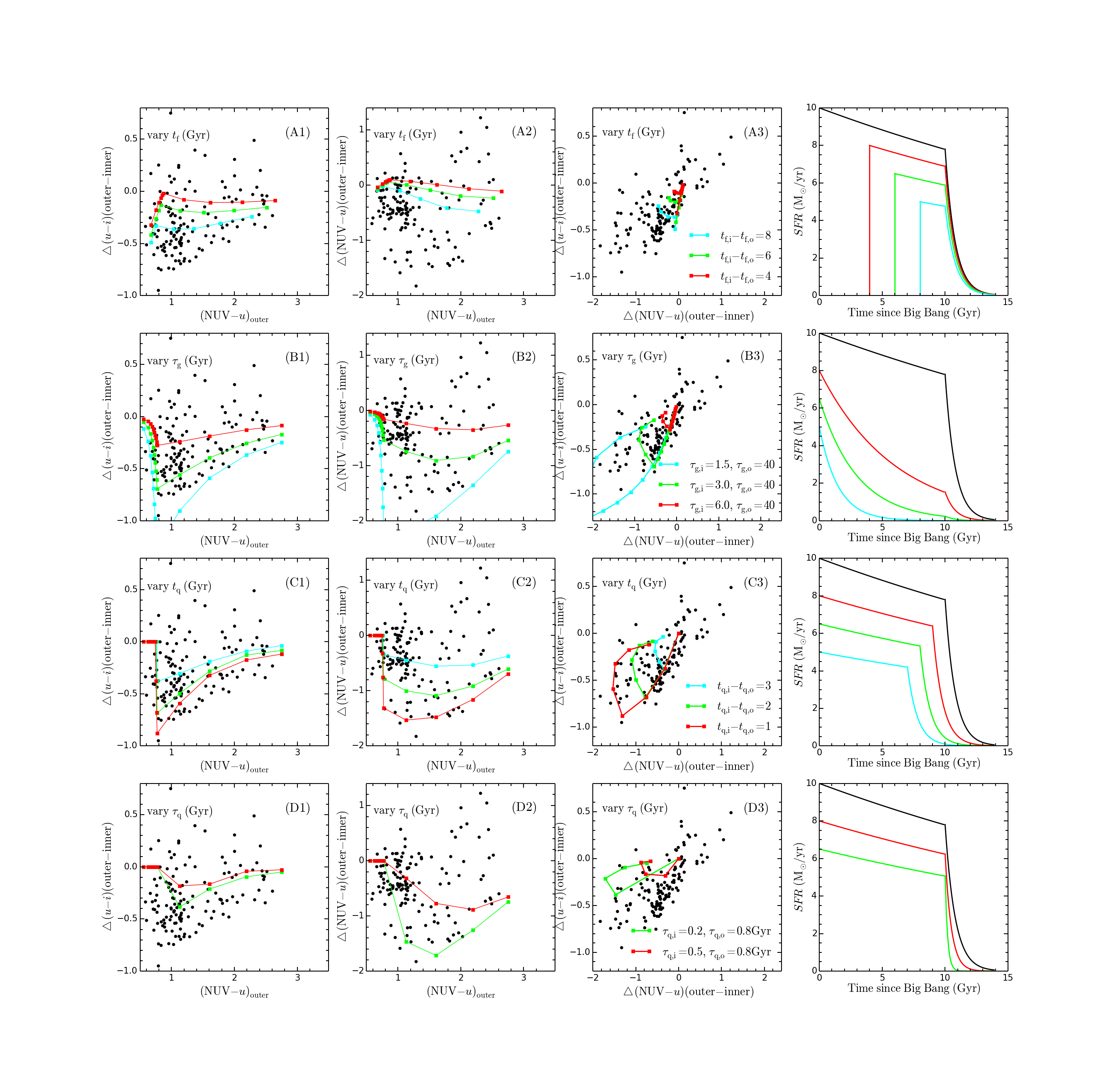}
\caption{$u-i$ and NUV$-u$ colour gradients of local galaxies.
Data points are identical in each row but overlaid with models of different `gradient parameter'.
{The difference between the inner and outer parts in the gradient parameter is shown in the legend in the third column.}
First row: model A, outer parts form at later $t_{\rm f}$. Second row: model B, inner parts have smaller $\tau_{\rm g}$ with faster
evolution speed  {during the} growth stage. Third row: model C, inner parts start quenching at an earlier $t_{\rm q}$. Bottom row: model D,
inner parts have smaller $\tau_{\rm q}$ with faster evolution speed  {during the} quenching stage.
{The fourth column shows the SFH adopted for the corresponding model in each row. The black line represents the SFH adopted for the outer  {region} except for the first row where the black line indicates the inner  {region}. The {coloured} lines indicate the other  {region} of the model galaxy
with various choices for the gradient parameter.}
}
\label{figure5}
\end{figure*}

{\sl Formation time $t_{\rm f}$:}
In the first row of {Figure 5}, we explore the colour gradient  {produced by different formation time $t_{\rm f}$ between the inner and outer regions, i.e., the gradient parameter in this model is the formation time. To produce the negative colour gradient, the inner part is required to form at an earlier time than the outer part.}
This is the first type of an `inside-out growth' model  {named `model A'}.
{In this model,}
the gradient in $u-i$  {colour} reaches the maximum  {at the beginning} when the outer part just forms and then  {monotonically} decreases to zero  {when} the outer part  {becomes} quiescent  {with NUV$-u>2.7$}. 
The outer part is  {assumed} to quench simultaneously with the inner part,  {suggesting} a shorter growth phase  {due to a later formation}.
{It can be seen that} model A only produces slight or mild colour gradients,
and {fails} to explain  {the galaxies} with  {steep} gradients,  {especially in NUV$-u$ colour}.
Changing the other  {free} parameters does not improve the fitting significantly.
Although  {model A cannot be} totally
{ruled} out as an explanation for {the mild colour gradients observed in some} galaxies,
we  {conclude} that the observed colour gradients are not  {mainly driven by differing formation times between the inner and outer regions}.
  
{\sl Growth e-folding time $\tau_{\rm g}$:}
In the second row  {of Figure 5}, we present the model with  {varying} e-folding time {at the growth phase $\tau_{\rm g}$, named} 
`model B'. In this model, the relatively redder colour of the inner part is due to its relative faster evolution {during the growth phase}. 
This is  {the second type} of `inside-out growth' model.
{The model} colour gradient  {steepens during the growth} stage and  {peaks} when  {the} quenching  {phase} starts. 
{The} e-folding times of the outer part  {during the growth phase ($\tau_{\rm g,o}$) is set} to be 40 Gyr.
This value is taken from \citet{noeske2007} who derived the SFH for star-forming galaxies as a function 
of stellar mass by  {studying} the cosmic evolution of mass-SFR (i.e. main-sequence) relation.
{As our galaxy sample has a median mass of $10^{9.92\pm0.28} M_{\odot}$, we adopt the SFH of galaxies with stellar mass of $10^{10}M_{\odot}$ in \citet{noeske2007} which have an e-folding time of 40 Gyr.}
Model B  {can reproduce the observed} negative colour gradients fairly well in both NUV$-u$ and $u-i$. 
{Given our choices of other parameters, we find} the inner part 
 {needs an} e-folding time  {during the growth phase}
$\tau_{\rm g,i}$ ranging in [3, 6] Gyr.

{\sl Quenching start time $t_{\rm q}$:}
In the third row of  {Figure 5}, the `gradient parameter' is the quenching start time $t_{\rm q}$.
{In this scenario, to match the negative colour gradient, the inner part of galaxies needs to start quenching earlier than the outer part.}
This is our first  {type of} `inside-out quenching' model,  {named `model C'.}
Before  {the inner part starts} quenching, the colour gradients are zero  {because the SFH} for the inner and outer parts are  {exactly} identical.
When the inner part starts quenching while the outer part is still in  {the} growth stage, a  {dramatic} negative gradient 
in NUV$-u$  {colour} and  {a mild negative} gradient in $u-i$  {colour develop within two or three Gyrs and peak at the moment when the outer region starts quenching}. {After that, the negative colour gradients begin to decrease until the outer part becomes quiescent.}
{Like model B,} model C also  {covers} the observed  {negative} colour gradients very well.
 
{\sl Quenching e-folding time $\tau_{\rm q}$:}
In the bottom row, we show the model  {with varying} quenching e-folding time $\tau_{\rm q}$,
called `model D'.
{To produce negative colour gradient, the galaxy inner part {needs} to have a shorter e-folding time during the quenching phase.}
This is  {the} second  {type} of `inside-out quenching' model.
The quenching e-folding time of the outer part is 
 {set} to be 0.8 Gyr. {According to the findings in Paper I that, the typical quenching time  {scales} of local galaxies should be
within the range of [0.2,1] Gyr.} 
{Changing the quenching time scales within [0.2, 1] Gyr does not change our results significantly.} 
Comparing to the  {observations},  {the colour gradient produced by model D is too steep in NUV$-u$ or too shallow in $u-i$ and
deviates} from the main distribution of galaxies as shown in panel D3 in  {Figure 5}.
{This mismatch cannot be improved by varying the other free parameters.}
Therefore,  {we conclude that} the observed colour gradient  {cannot be driven} by different e-folding time  {during the quenching phase between} the inner 
and outer parts of galaxies.

To summarize, we find two models,  {one inside-out growth model (model B)} with  {a} gradient in  {the growth}
e-folding time and  {one inside-out quenching model (model C)} with  {a} gradient in  {the} quenching start time, 
 {are able to reproduce the ranges of observed negative} colour gradients  {in NUV$-u$ and $u-i$}.
For convenience in the following discussion, we refer to our 
model B as the `inside-out growth' model and model C as the `inside-out quenching' model.
The other two models with  {gradients} in the other two parameters  {of} the two-phase SFH  {fail to explain the observation and are} ruled out. This result shows the power of colour gradients in NUV$-u$ and $u-i$ colour in differentiating variations  {of} 
SFH across galaxies.
However, the `inside-out growth' and `inside-out quenching'  {models cannot} be  {distinguished} only by  {the} colour gradients. 
Other independent {constraints are} needed. 

\section{Inside-out growth or inside-out quenching?}
\begin{figure*}
	\includegraphics[width=17cm,viewport=90 60 660 270,clip]{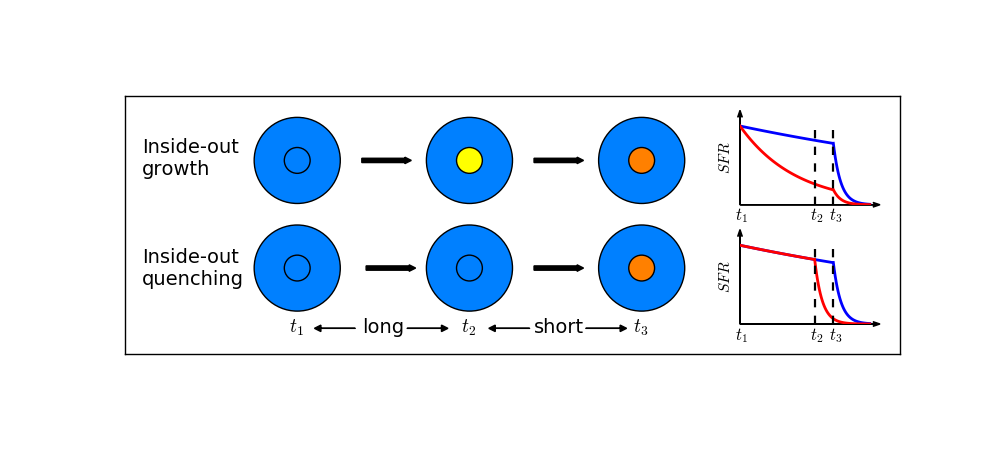}
	\caption{Schematic cartoon that shows the different colour gradient evolution history between the inside-out growth and inside-out quenching model. {The inside-out growth model invokes different e-folding time at the growth phase $\tau_{\rm g}$ between te inner and outer parts of galaxies while the inside-out quenching model has a different quenching time $t_{\rm q}$ (See Figure 5 and the text for more detailed discussion). }
	The star formation history of the two scenarios are also shown at the right side. Blue and red lines represent the outer region and inner region, respectively.}
	\label{figure6}
\end{figure*}

{Figure 6 shows a schematic cartoon to illustrate the {buildup} history of the colour gradient and SFH for the inside-out growth and inside-out quenching models. The cartoon shows the colour of the inner and outer parts of {two mock galaxies} at three points in time.
When the galaxy is just formed, no colour gradient exists, and both the inner and outer parts are blue. Later, the inner part of the mock galaxy with inside-out growth becomes redder than the outer part because the SFR of the inner part declines slightly faster. In contrast, the model galaxy with inside-out quenching still has zero colour gradient until the inner part start to quench. However, in the latter model, shortly after the inner part starts to quench at the time point $t_2$, it becomes much redder than the outer part rapidly.}

As already shown in  {Figure 5, the remaining `inside-out growth' and `inside-out quenching'} models can explain the observed colour gradients
by only varying the `gradient parameter'. {However, the gradients in these models show up at different evolutionary stages and last for different amounts of time. Thus, the two models should have significantly different predictions for the colour gradient distribution among galaxies.}
{Previously, in Sec 4 and Fig 5, we were only concerned with the range of the gradient values, rather than its exact distribution. Thus, the gradient distribution provides an additional, independent constraint.}
To differentiate  {the two models},
we utilize the observed  {colour gradient distribution} and compare it with the predictions of the models.

The left panel of Figure 7 shows the colour gradient in $u-i$ as a function of NUV$-u$ colour of the outer part.  
Only galaxies with negative colour gradient in both NUV$-u$ and $u-i$ are  {included in the plot}. 
The green lines at NUV$-u=1.3$ and NUV$-u=2.7$ are used to divide the NUV$-u$ space into three regions, star-forming, transition, and quiescent regions from the left to the right. The dividing boundary is adopted according to the number density drops in the NUV$-u$ colour distribution.
{The predictions of a typical `inside-out growth' and `inside-out quenching' models are shown in cyan and orange colour, respectively.
The parameter adopted for the two models are listed in Table 1. 
As expected, the two models cover the observed colour gradient ranges equally well but develop the colour gradient differently.}
In the inside-out growth model, the  {negative} colour gradient gradually increases during the growth stage 
while in the inside-out quenching model the colour gradient  {increases dramatically} after the inner part starts quenching and 
reaches its peak in a short time of 3 Gyr.

Due to this fast evolution of colour gradient, the inside-out quenching model cannot fit the colour gradient distribution of galaxies very well. 
{On one hand, the inside-out quenching model suggests zero colour gradient for the first several Gyrs after galaxy formation.}
{Assuming galaxies form with a constant rate at different epochs of the universe and follow the same star-formation history, in the inside-out quenching scenario,}
we should find many galaxies with zero colour gradients which, however, are not present in observation. 
{It is hard to imagine we are catching all galaxies during the brief moment of quenching (up to 3 Gyrs at most) and they are all doing it simultaneously.}
On the other hand, judging from the age marks on the inside-out quenching model track, the model lasts for a similar time period with negative colour gradient at the star-forming and transition regions. With a constant galaxy formation rate, we should find similar numbers of galaxies in the two regions. However, by doing a simple counting, it is easy to find more galaxies with negative colour gradient located in the star-forming region compared to those in the transition region.

We demonstrate the above point quantitatively {by comparing the observations and predictions of each model in the colour gradient distribution. Since the comparison is similar in the $u-i$ and NUV$-u$ colour gradient, for simplicity, we only discuss the comparison in the $u-i$ colour gradient.} 
{In each model, assuming} a constant galaxy formation rate, we can compute how many galaxies we expect to observe with a certain colour gradient. When comparing to the data, we do the comparison for galaxies in star-forming region (defined by NUV-$u_{\rm couter} < 1.3$) and transition region (defined by $1.3 < {\rm NUV}-u < 2.7$) separately. 
The colour gradient distribution predicted by the models are derived by assuming a constant galaxy formation rate and counting the model points in each $u-i$ colour gradient bin. To be consistent with the data, the colour gradient distribution of models are also separated according to the same colour cuts, i.e. model points with (NUV$-u)_{\rm outer}<1.3$ are counted in the `star-forming' panel and vice versa. Therefore this comparison is not dependent on the choice of colour cut. 
The middle and right panels of Figure 7 shows the comparison in $u-i$ colour gradient distribution for the galaxies in star-forming and transition regions, respectively. The black dashed histograms represent the observed colour gradient {distribution}; cyan histograms are for the colour gradient distribution predicted by the inside-out growth model; orange histograms are for the distribution predicted by the inside-out quenching model.
{The distribution of the models can be scaled up and down by adopting {different time grids}. We adopt a time grid with 1 Gyr spacing for the models shown in the left panel but a time grid with 0.15 Gyr spacing for the models in the middle and right panels. The denser time grid (0.15 Gyr) is needed to roughly match the total number of observed galaxies.}



The comparison confirms our previous interpretation that the `inside-out quenching' scenario would yield an $u-i$ colour gradient distribution that does not match the observed distribution well. 
In contrast, the inside-out growth model predicts a flat colour gradient distribution in the star-forming region which is consistent with the observation. Moreover, this model also reproduces more galaxies with negative colour gradient observed in star-forming than transition region. 
Therefore, based on the $u-i$ colour gradient distribution, the inside-out growth scenario is preferred to the inside-out quenching scenario. 
{For star-forming galaxies (middle panel of Fig 7), the KS probabilities are 0.02 for the inside-out growth model and 1.35e-11 for the inside-out quenching model. For the transition galaxies, the KS probabilities are 0.74 for the inside-out growth model and 0.12 for the inside-out quenching model. {This suggests that} the inside-out quenching model is less favoured by the observations. The KS probability can only be considered in the relative comparison. And we should not take the absolute value too seriously as the model is overly simplistic.}
This analysis is limited by the sample size and more observations in spatially-resolved UV imaging or {spectroscopy} are needed to increase the significance of the result.

The only way to resurrect the inside-out quenching model is to assume all galaxies formed at roughly the same time and they are all evolving synchronously. 
Based on the deep imaging and spectroscopic survey, the cosmic SFH is found to peak at $z\sim2$ ($\sim3.5$ Gyr after big bang, \citealt{madau2014}) and decline significantly by an order of magnitude from the peak to now. 
However, the peak in the cosmic SFH at $z\sim2$ should not be considered as an evidence for a peak formation of galaxies observed today. This is because the progenitor of today's star-forming galaxies may be too faint or even not yet formed to be counted in the cosmic SFH which is obtained by deep surveys at $z\sim2$ with typical mass completeness above $10^{9.5} {\rm M_{\odot}}$ \citep{madau2014}.

Regarding the assumption about a common SFH among all galaxies, {although} the inside-out growth scenario may be representative for most of the local star-forming galaxies, not all galaxies necessarily follow this SFH. As we mentioned in \textsection4.1, many galaxies have blue star-forming outer parts but have red quiescent inner parts. These galaxies can also be identified in Figure 5 with blue outer part ((NUV$-u)_{\rm outer}<1.3$) and distinctively steep NUV$-u$ colour gradient ($\triangle ({\rm NUV}-u)_{\rm outer-inner}<-1$). In the second row with inside-out growth model B, the extremely steep NUV$-u$ colour gradient requires a very short e-folding time of 1.5 Gyr for the inner part, which is close to the e-folding time of 0.8 Gyr for the quenching phase. In these cases, the difference between the inside-out growth and inside-out quenching model becomes ambiguous. 
An inside-out growth model with extremely short e-folding time for the inner part could also be considered to be an inside-out quenching model. 
Since galaxies with such extreme colour gradient are rare and are only found in the left tail of the colour gradient distribution in Figure 7,
the comparison is not significantly affected by the potential deviation of SFH in these galaxies. 
Recently, \citet{tacchella2015} studied a sample of massive star-forming galaxies at $z\sim2.2$ with deep imaging and integral field unit (IFU) observation. They found evidence for inside-out quenching in the most massive objects ($\sim10^{11}{\rm M_{\odot}}$) which show similar surface mass density profiles to the local quiescent galaxies of the same mass range in the galaxy center. Limited by the galaxy sample size, we are not able to further explore whether the preference of inside-out growth and inside-out quenching scenario is dependent on the stellar mass of galaxies. 
To significantly improve the spatially-resolved sample, a high spatial resolution and wide-field UV imaging survey is needed.

\begin{figure*}
\includegraphics[width=17.4cm,viewport=15 0 1047 345,clip]{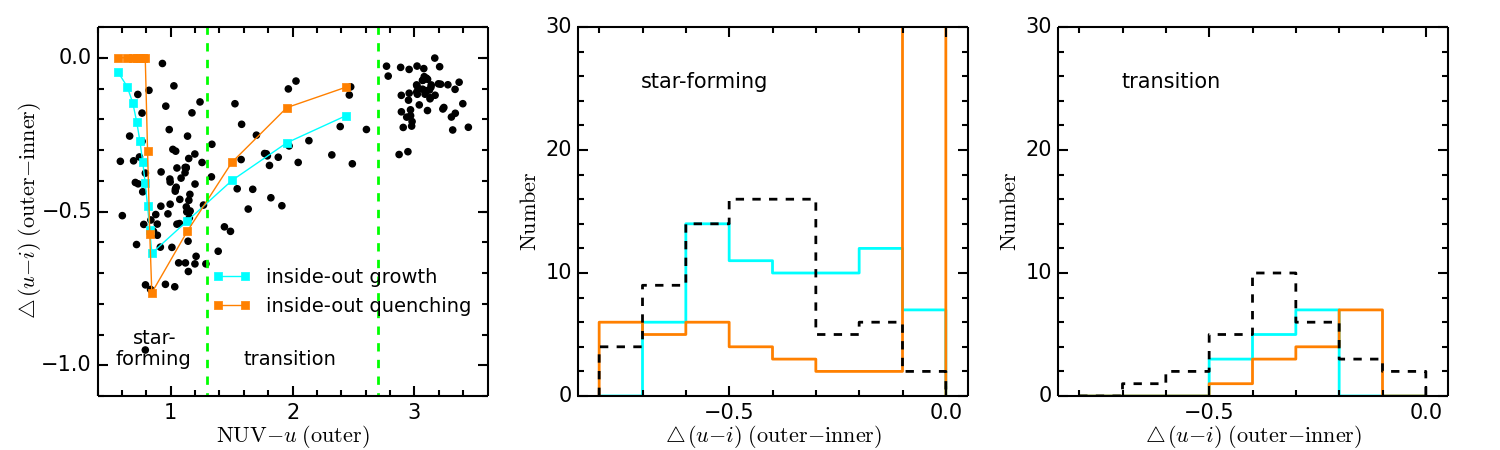}
\caption{ Left: Colour gradient in $u-i$ as a function of NUV$-u$ colour of outer regions.
{Cyan} and  {orange} curves represent the adopted `inside-out growth' and 
`inside-out quenching' models, respectively. Green dashed lines indicate the boundary for separating  galaxies during the star-forming or transition stages. 
Middle and right: $u-i$ colour gradient distribution for galaxies and models with (NUV$-u)_{\rm outer}<1.3$ (middle) and $1.3<({\rm NUV-}u)_{\rm outer}<2.7$ (right). Black dashed histograms represent the observed colour gradient distribution; cyan histograms are for the colour gradient distribution predicted by inside-out growth model; orange histograms are for the distribution of inside-out quenching model.}
\label{Figure7}
\end{figure*}

\begin{table*}
  \tabcolsep=0.3cm
  \centering
  \caption{Parameters adopted for the `inside-out growth' and `inside-out quenching' models in Figure 7.}
  \label{table1}
  \begin{tabular}{lcccccc}
  
\hline\hline
 \multicolumn{1}{c}{model} & $\tau_{\rm g,i}$ & $\tau_{\rm g,o}$ & $t_{\rm q,o}$ & $\triangle t_{\rm q}$ & 
 $\tau_{\rm q}$ & $E(B-V)$  \\
        & Gyr & Gyr & Gyr & Gyr & Gyr & \\
\hline
  inside-out growth & 3 & 10 & 10 & - & 1 & 0.1 \\ 
  inside-out quenching & - & 10 & 10 & 3 & 1 & 0.1 \\
\hline
  \end{tabular}\\
\end{table*}

\section{summary}
In this work, we investigate the internal variation of SFH across galaxies by  {analyzing} the 
colour gradients in NUV$-u$ and $u-i$ for a local spatially-resolved galaxy sample. We 
 {study the inner ($r<0.3r_{\rm e}$) and the outer ($0.5{\rm r_e}<r<{\rm r_e}$) regions in each galaxy.}
The  {colour distribution} 
of  {these two regions} show a systematic offset.
{Generally, the outer parts of galaxies are bluer compared to the inner parts, confirming the} negative colour gradient 
{found in the local galaxies.}

{We first ruled out dust extinction gradient or metallicity gradients as the dominant cause of the colour gradients. To further explore the origin of the observed negative colour gradient, we use stellar population synthesis model and investigate the colour gradient predicted by radially-varying SFHs in galaxies. We adopt a two-phase SFH which consists of a growth phase with long timescale and a subsequent quenching phase with short timescale.} 
{Compared to the observation, we find two simple scenarios  {can cover} the  {observed} negative colour gradient  {ranges} in NUV$-u$ and $u-i$ equally well. In the first scenario, the inner regions of galaxies tend to evolve faster during the growth phase with a shorter timescale compared to the outer regions and therefore the inside regions turn out to be redder. We call this model `inside-out growth' model. In the other plausible scenario, the inner part of galaxies are redder because they cease their star formation (i.e. quench) earlier than the outer part. This model is called `inside-out' quenching model.} 

To differentiate the inside-out growth and inside-out quenching scenario, we  {compare the observation with predictions of the two models in the $u-i$ colour gradient distribution. Since the colour gradients develop differently in the two scenarios, the colour gradient distribution predicted by the two models are therefore different.
With assumptions of constant galaxy formation rate and  {a common SFH followed by most galaxies}, the inside-out quenching model cannot  {reproduce} the observed  {colour gradient} distribution. In contrast, the  {colour gradient distribution predicted by the} inside-out growth model {is in good agreement with the observed distribution.}
Therefore,  {we conclude that} the inside-out growth scenario is {supported} by the observed colour gradients of galaxies in NUV$-u$ and $u-i$.
Nevertheless, we can not rule out the possibility that a minor portion of galaxies, especially the massive ones \citep{tacchella2015}, may experienced an inside-out quenching SFH. Moreover, the significance of the result is limited by the sample size and more
spatially resolved UV observations are needed for a higher significance study.}
{Besides}, we urge all discussions on spatially-resolved evolution studies to clearly define what is meant when 
speaking of `inside-out growth' or `inside-out quenching'.

\section*{Acknowledgements}
{\sl Galaxy Evolution Explorer} (GALEX) is a NASA Small Explorer, launched in 2003 April. We gratefully acknowledge NASA’s support for the construction, operation and scientific analysis for the GALEX mission, developed in cooperation with the Centre National d’Etudes Spatiales of France and the Korean Ministry of Science and Technology.

Funding for the SDSS and SDSS-II has been provided by the Alfred P. Sloan Foundation, the Participating Institutions, the National Science Foundation, the U.S. Department of Energy, NASA, the Japanese Monbukagakusho, the Max Planck Society, and the Higher Education Funding Council for England. The SDSS Web Site is http://www.sdss.org/.

The SDSS is managed by the Astrophysical Research Consortium for the Participating Institutions. The Participating Institutions are the American Museum of Natural History, the Astrophysical Institute Potsdam, the University of Basel, the University of Cambridge, the Case Western Reserve University, the University of Chicago, Drexel University, Fermi- lab, the Institute for Advanced Study, the Japan Participation Group, Johns Hopkins University, the Joint Institute for Nuclear Astrophysics, the Kavli Institute for Particle Astrophysics and Cosmology, the Korean Scientist Group, the Chinese Academy of Sciences (LAMOST), the Los Alamos National Laboratory, the Max-Planck-Institute for Astronomy (MPIA), the Max-Planck-Institute for Astrophysics (MPA), New Mexico State University, Ohio State University, the University of Pittsburgh, the University of Portsmouth, Princeton University, the United States Naval Observatory, and the University of Washington.

\end{document}